\documentclass[9pt,twocolumn,twoside]{opti_preprint}

\setboolean{shortarticle}{false}

\usepackage{soul}

\title{Transmission matrix approaches for non-linear fluorescence excitation through multiple scattering media}

\author[1,*]{Mickael Mounaix}
\author[1]{Duc Minh Ta}
\author[1]{Sylvain Gigan}

\affil[1]{Laboratoire Kastler Brossel, ENS-Université PSL, CNRS, Sorbonne Universit\'{e}, Coll\`{e}ge de France, 24 rue Lhomond, 75005 Paris, France }

\affil[*]{Corresponding author: mickael.mounaix@lkb.ens.fr}

\begin{abstract}
Several matrix approaches were developed to control light propagation through multiple scattering media under illumination of ultrashort pulses of light. These matrices can be recorded either with spectral or temporal resolution. Thanks to wavefront shaping, temporal and spatial refocusing have been demonstrated. In this work, we study how these different methods can be exploited to enhance a two-photon excitation fluorescence process. We first compare the different techniques on  micrometer-size isolated fluorescent beads. We then demonstrate point-scanning imaging of such fluorescent microbeads located after a thick scattering medium, at a depth where conventional imaging would be impossible because of scattering effects. 
\end{abstract}

\begin{document}

\maketitle
\pagestyle{plain}
\ifthenelse{\boolean{shortarticle}}{\abscontent}{}

Multiple scattering of coherent light is usually considered detrimental for imaging, as information transported by light seems utterly randomized. Nonetheless, the obtained speckle pattern can be controlled with a spatial light modulator (SLM) via wavefront shaping~\cite{rotter_light_2017,mosk_controlling_2012,Vellekoop:15}. Among the various methods to control  the intensity of light on a sensor (a CCD camera for instance) using a SLM~\cite{Vellekoop2007,Cui:10}, a particularly interesting  one is to record the optical transmission matrix, relating the output field on the spatially-resolved sensor to the input field on each of the SLM pixels~\cite{popoff_measuring_2010}. 

Scattering of broad spectrum of light can induce spatio-temporal distortions of the transmitted light, known as spatio-temporal speckle. In essence, different wavelengths may result in different speckle patterns, if the medium is too thick~\cite{andreoli_deterministic_2015}. For an ultrashort pulse, it leads to a temporal broadening, in addition to a drastic loss of fluence. Consequently, all non-linear processes relying on the ultrashort duration of a pulse of light are then strongly inhibited. In particular, two-photon excitation fluorescence processes (2PEF)~\cite{zipfel2003nonlinear}, commonly used for deep imaging and microscopy~\cite{helmchen2005deep}, are strongly limited by scattering. 

Various approaches were proposed to control light transmission to achieve spatio-temporal focusing, which consists in concentrating the output pulse at a given spatial position, while ensuring that the achieved output pulse retrieves almost its initial duration. They are mostly based on the use of  iterative optimization algorithms~\cite{katz_focusing_2011,aulbach_control_2011,aulbach2012spatiotemporal}, digital optical phase conjugation~\cite{morales2015delivery}, or on the measurement of different kinds of transmission matrices (TM). On the one hand, the Multi-Spectral TM (MSTM)~\cite{mounaix_spatiotemporal_2016} is a stack of monochromatic TMs, measured with a proper sampling of the spectral components of the ultrashort pulse. On the other hand, the Time-Resolved TM (TRTM)~\cite{mounaix_deterministic_2016} is a tensor, made of time-gated TMs, measured for a set of different arrival times of the output pulse. These two stacks of TMs rely on the use of an external reference beam during their measurement processes. In contrast, the Broadband TM (BBTM)~\cite{Mounaix:17} is a single operator measured with a self-referencing beam for all spectral components simultaneously. Although it cannot achieve spatio-temporal focusing by essence of its measurement process, it leads to two-fold spectral/temporal~\cite{Mounaix:17} compression, and to interesting polarization effects~\cite{de2017polarization}, which could lead to potential applications in non-linear imaging through biological systems via its fast and simple measurement. Other methods based on time-gated reflection matrices have also been proposed with similar results~\cite{choi_measurement_2013}.

In this letter, we  compare several transmission matrix techniques to enhance and to control a 2PEF process through a multiple scattering medium. In particular, we show that a MSTM allow precise tuning of the 2PEF via control of second order pulse dispersion.  We then perform point-scanning imaging through the same static scattering sample, as a new route towards non-linear imaging through disordered systems.

\begin{figure}
\centering
\includegraphics[width=\linewidth]{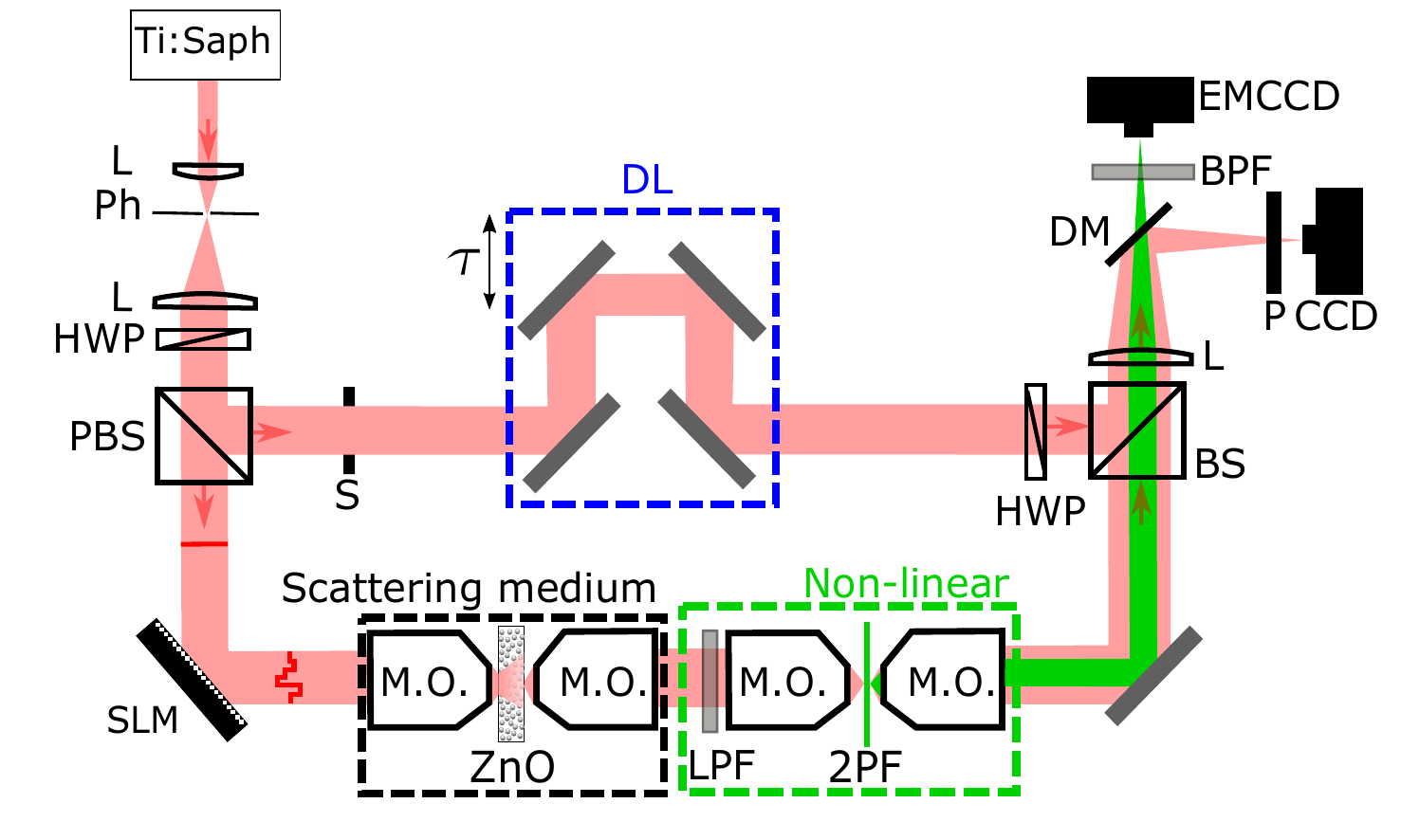}
\caption{Experimental setup. (L): lens; (Ph): pinhole; (HWP): half wave plate; (PBS): polarized beam splitter; (S): shutter; (DL): delay line; (SLM): spatial light modulator; (MO): microscope objective; (LPF): long pass filter; (DM): dichroic mirror (BPF): band pass filter; (P): polarizer}
\label{fig:exp_setup}
\end{figure}

Fig.~\ref{fig:exp_setup} illustrates the experimental setup. A Ti:Saph laser source (MaiTai, Spectra-Physics) generates ultrashort pulses of duration $\sim$ 100~fs centered around 800~nm. The pulse is split with a polarized beam splitter (PBS) into a reference pulse and a controlled pulse. The wavefront of the control pulse is modulated with a phase-only SLM (LCOS-SLM X10468-2, Hamamatsu), before it propagates through a thick layer (thickness $\sim~100~\mu$m) of ZnO nanoparticles (polydisperse diameter < 5~$\mu$m, Sigma-Aldrich) via a microscope objective (MO). The output plane of the scattering medium is imaged on fluorescent microbeads (diameter 1 $\mu$m, Thermo Fisher) immersed in an aqueous gel (Sigma-Aldrich), via a MO (Olympus, MplanFL N, 100x, NA 0.85).  The microbeads are used here as a proof of principle experiment. The protocol and the achieved resolution would be similar with fluorescent cells, as they rely on linear measurements. Only speed would depend on the amount of the recorded fluorescent signal. A long-pass filter (LPF) is placed between the two pairs of MO to filter out residual autofluorescence from the scattering medium. Transmitted light (both linear signal at 800 nm and fluorescence) is collected with an oil-immersed MO (Nikon Plan Apo VC, 60x, NA 1.4). The output light is overlapped on a beam-splitter with the reference beam, whose delay is controlled with a delay line. A shutter (S) allows blocking the reference beam when needed. A tube lens and a dichroic mirror (DM) allows imaging the linear output intensity on a CCD camera (Manta, AVT) and the fluorescence signal on an EMCCD camera (iXon Ultra, Andor). 

The different TMs of the scattering medium, as listed above, can all be measured with the experimental setup of Fig.~\ref{fig:exp_setup}, by choosing appropriately either broadband or CW light, and using or not the reference arm. Prior to measuring the matrices, the spectral correlation bandwidth of the scattering medium $\delta \lambda_m$ is measured~\cite{andreoli_deterministic_2015}. It corresponds to the minimum distance in input wavelengths that corresponds to two uncorrelated speckle patterns, and inversely related to the average traversal time  $\tau_m$ of light in the material. $\tau_m$ is closely related to the diffusion time of light through the sample~\cite{Vellekoop05}. 
The time-of-flight (ToF) distribution of the scattering medium can be measured by scanning the delay line, and recording the interferogram as function of delay $\tau$ with an Interferometric Cross-Correlation technique (ICC)~\cite{mounaix_spatiotemporal_2016}. The average confinement time of photons $\tau_m \sim 1$~ps can be determined from the exponential decay of the ToF distribution. 

For all cases, the measurement of a single TM consists in displaying a series of $N_{\text{SLM}}=1024$ Hadamard patterns on the SLM, and measuring the corresponding transmitted field on the CCD camera using phase-shifting holography~\cite{popoff_measuring_2010} with a reference beam, either co-propagation (BBTM) or from the reference arm (TRTM and MSTM). (a) For the MSTM or for a monochromatic TM, the Ti:Saph is mode unlocked and used as a tunable CW source. Measuring the MSTM consists in measuring a set of $N_\lambda = \Delta \lambda / \delta \lambda_m \simeq 11$ monochromatic TM, where $\Delta \lambda \simeq 12$~nm stands for the spectral bandwidth  of the ultrashort pulse at FWHM;
(b)  For the TRTM, we only measure a single time-gated matrix from the full TRTM of the scattering medium. The laser is used in pulsed mode, and we perform gated measurement using phase-shifting holography with the low-coherence reference pulse, setting the delay line to the maximum of the ToF distribution;
(c) The BBTM of the scattering medium is measured with the protocol developed in~\cite{Mounaix:17}, essentially as in \cite{popoff_measuring_2010} but with a broadband pulse. The shutter is closed, and we use a non-modulated part on the the SLM as a reference beam. After measuring the MSTM, the laser is turned back to mode-locked pulsed operation, and the shutter S is closed, for all focusing and fluorescent experiments.

\begin{figure} [b!]
\centering
\includegraphics[width=\linewidth]{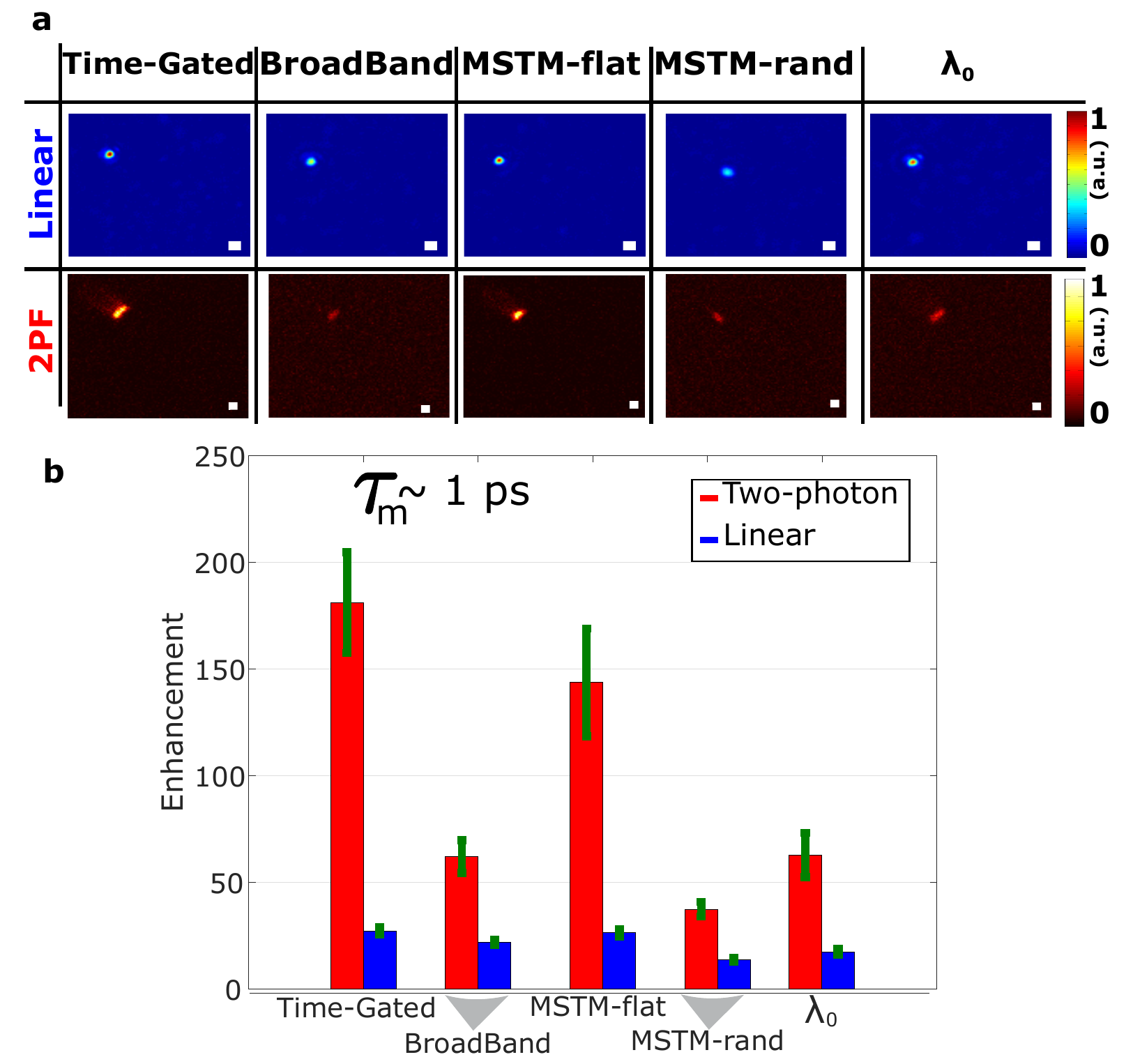}
\caption{Comparison of 2PEF signal between the different TM approaches of the multiple scattering sample: the time-gated TM measured at the maximum of the time-of-flight distribution, the Broadband TM, and three solutions using the Multi-Spectral TM : respectively a flat spectral phase MSTM-flat ensuring a short pulse, and two solutions ensuring spatial  but no temporal focusing, namely a random spectral phase MSTM-rand and exploiting only the central wavelength TM at $\lambda_0$. (a) Top line: linear signal of focusing on a single spatial position exploiting the different TM. Bottom line: corresponding 2PEF signal. Intensity normalized by focus intensity obtained with the time-gated TM. Scale bars: $2 \mu$m. (b) Linear (blue) and 2PEF (red) signal-to-background ratios (SBR) of focusing, averaged over 9 different foci. Green lines stand for the standard deviation of the SBR.}
\label{fig:2PEF_comparison}
\end{figure}

Transmitted light can be focused on a given target by phase conjugation, i.e. exploiting the transpose conjugate of the TM~\cite{popoff_measuring_2010}. In the following, we focus light on a single isolated fluorescent microbead. In addition to focusing light with the time-gated TM and the BBTM, three different experiments are carried out with the measured MSTM. It consists in focusing only the central wavelength ($\lambda_0 = 800$~nm) of the pulse, via phase-conjugating the corresponding monochromatic TM of the MSTM, or focusing all the spectral components and simultaneously adjusting their corresponding spectral phase relationships. Two configurations are studied: imposing either a flat spectral phase (MSTM-flat) leading to spatio-temporal focusing, or a deliberated random spectral phase (MSTM-rand) distribution which ends to spatial-only focusing~\cite{mounaix_spatiotemporal_2016}. 

Fig.~\ref{fig:2PEF_comparison}a shows the experimental results of both linear (at $\sim$ 800 nm) and 2PEF transmitted images, measured respectively with the CCD camera (exposure time $T_e^{\text{CCD}}~=~$10~ms) and the EMCCD camera (exposure time $T_e^{\text{EMCCD}}~=~$10~s, electronic gain 1000). The signal-to-background ratio (SBR) is calculated as the ratio of the focus intensity over the spatially-averaged background intensity.
One can  first notice that all the linear SBR of the various foci take similar values of $15-25$. Indeed, the linear SBR depends only on the average power, which scales with the number of controlled SLM pixels, and not on the peak power (i.e. the temporal properties). The SBR of the 2PEF is always higher, but cannot be directly related to the linear enhancement, probably due to spurious background signal. However, we observe clearly a strong 2PEF focus in all cases, and the 2PEF SBR depends strongly on the temporal duration of the transmitted pulse. On the one hand, both the time-gated TM and the MSTM-flat focusing leads to spatio-temporal focusing: the achieved output pulse should get close its initial Fourier-limited duration~\cite{mounaix_spatiotemporal_2016,mounaix_deterministic_2016}, leading to $>130$ SBR. On the other hand the BBTM, the monochromatic focusing and the MSTM-rand lead to spatial-only focusing (with only moderate compression for the BBTM): the output pulse remains temporally elongated, leading to a  comparatively lower 2PEF compared to a spatio-temporal focus.
The average SBR of focusing over 9 different fluorescent microbeads is illustrated in Fig.~\ref{fig:2PEF_comparison}b for the five different cases. One can observe that the time-gated focusing has a slightly higher SBR than the MSTM-flat. Indeed, the time-gated focusing leads to almost perfect spatio-temporal focusing as all the spatial degrees of freedom of the SLM are exploited to enhance light intensity at a time fixed by the low-coherence gate. When measuring the MSTM,  errors on the spectral phase estimation in particular due to long measurement times, as well as the finite spectral sampling, lead to a comparatively lower temporal enhancement. 

\begin{figure}[t!]
\centering
\includegraphics[width=\linewidth]{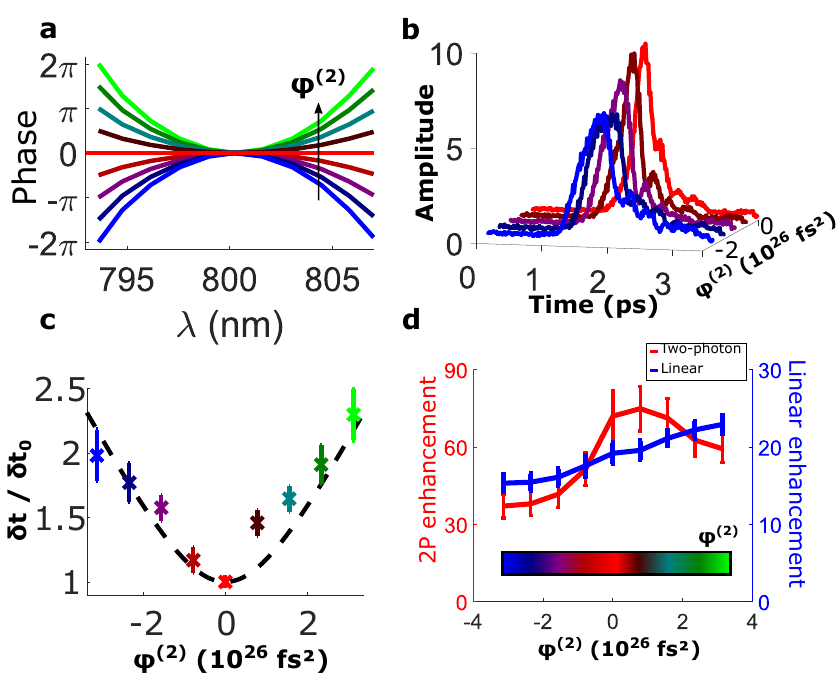}
\caption{Controlled chirped pulse with the Multi-Spectral TM. (a) Imposed spectral phase relationship $\varphi(\omega)$ between the different spectral components upon focusing with the Multi-Spectral TM: 4 positive chirps (green to dark green), 1 flat spectral phase (red) and 4 negative chirps (brown to blue) (b) Linear amplitude temporal profile retrieved via ICC for the 4 negative chirped pulses and the pulse with an imposed flat spectral phase relation. (c) Retrieved duration of the chirped pulse from ICC for the 9 different imposed chirp, compared to the predicted pulse duration (dashed line). (d) Linear and 2PEF SBR for the 9 chirped pulses. Error bars stand for the standard deviations over 9 different focus positions.}
\label{fig:MSTM_chirp}
\end{figure}

Beyond mere spatiotemporal focusing, the MSTM is particularly useful compared to the other approaches, as it provides additional spectral degrees of freedom to adjust the transmitted output pulse properties, in particular its spectral phase relationship $\varphi(\omega)$. In Fig.~\ref{fig:2PEF_comparison}, we only considered  a flat and a random $\varphi(\omega)$. While a spectral phase ramp enables to change the arrival time of the output pulse~\cite{mounaix_spatiotemporal_2016}, a quadratic spectral phase relationship  is well known to affect the temporal duration of the output pulse~\cite{Walmsley:09}. We demonstrate in Fig.~\ref{fig:MSTM_chirp} a controlled temporal broadening of the output pulse and its influence on the corresponding measured 2PEF signal. It is achieved by focusing the output pulse while simultaneously imposing a variable quadratic spectral phase. Nine different spectral curvatures profiles, that are shown in Fig.~\ref{fig:MSTM_chirp}a, are compared. For this purpose, the spectral phase relationship reads $\varphi(\omega) = \varphi^{(2)} (\omega-\omega_0)^2 /2$, with $\varphi^{(2)}$ the chirp and $\omega_0$ the central frequency of the pulse. For each chirp, the temporal profile at the focus position is measured with the ICC technique, and the process is repeated for 9 different focus positions. The averaged temporal profile enables then to determine the corresponding pulse width. Such averaged temporal profiles are shown in Fig.~\ref{fig:MSTM_chirp}b for the negative chirps (the positive chirps are not shown for clarity's sake). The  Fourier limited pulse duration $\delta t_0$ is measured by imposing a flat spectral phase relationship, and is estimated $\delta t_0 \sim$~200~fs in this experiment. The pulse duration $\delta t$ can be predicted theoretically, as it only depends on $\delta t_0$ and on $\varphi^{(2)}$. For a Gaussian ultrashort pulse, $\delta t$ reads~\cite{diels2006ultrashort}: $\delta t/\delta t_0 = \sqrt{(\delta t_0)^4 + 16 (\text{ln}(2)\varphi^{(2)})^2}/\delta t_0^2$.
In Fig.~\ref{fig:MSTM_chirp}c, we compare the measured pulse duration from the experimental temporal profiles, as shown in Fig.~\ref{fig:MSTM_chirp}b, with the theoretical expected duration calculated with the above formula. The two quantities are in good agreement. 
We then use these chirped focused pulses on fluorescent microbeads to excite 2PEF, and measure the corresponding linear SBR and 2PEF SBR. The experimental results for the 9 different chirped pulses are shown in Fig.~\ref{fig:MSTM_chirp}d. We note that the maximum 2PEF SBR is not measured for the flat spectral phase profile, but for a small positive chirp. Indeed, the imposed spectral phase relationship is applied on the CCD camera plane (See Fig.~\ref{fig:exp_setup}), where the TM was measured. However, the 2PEF sample is not located in the same plane: different optical elements add dispersion, in particular the fourth MO, which is expected to add a positive chirp~\cite{muller_dispersion_1998}. Consequently, the imposed flat spectral phase in the CCD plane is consistent with a slightly quadratic negative curvature in the 2PEF plane to compensate for this positive chirp. 
Note that although a more quantitative approach could be performed with the setup of Fig.~\ref{fig:exp_setup}, stability issues appeared when the experiment was performed. The experiment of Fig.~\ref{fig:MSTM_chirp}d was performed from positive chirp (green $\varphi^{(2)}$) to negative chirp (blue $\varphi^{(2)}$), i.e. from right to left on the curve, and the decrease with time of the linear SBR in Fig.~\ref{fig:MSTM_chirp}d is consistent with a slow decorrelation over time of the medium, which decreases the SBR values.

\begin{figure}
\centering
\includegraphics[width=\linewidth]{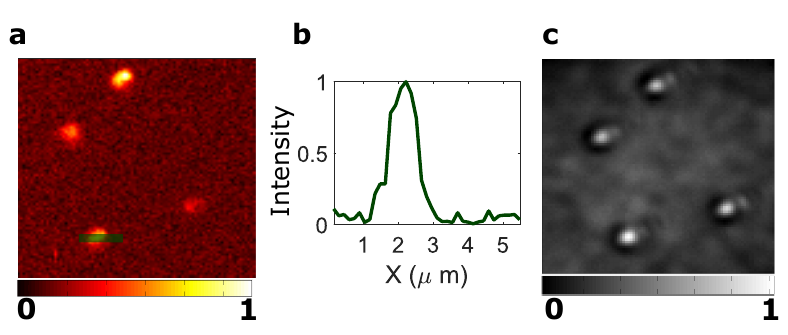}
\caption{Imaging fluorescent microbeads (ID $\phi = 1 \mu$m) with the time-gated TM. (a) Reconstructed image by scanning the focus using the time-gated TM and the SLM (80 $\times$ 80 foci), and collecting the total transmitted fluorescence for each focus position. (b) Intensity profile along 1 bead position (c) Mean linear speckle image, obtained by
averaging over 100 different random illumination patterns on the SLM. }
\label{fig:2PEF_imaging}
\end{figure}

Finally, we present an application of the TM measurement for point-scanning imaging fluorescent microbeads through a multiple scattering medium in Fig.~\ref{fig:2PEF_imaging}. For this purpose, we exploit the previously measured time-gated TM of the same scattering medium, as it enables reaching the highest 2PEF SBR, as we discussed in Fig.~\ref{fig:2PEF_comparison}. For each pixel on the CCD camera (RoI: $N_{\text{CCD}}~=$~80~$\times$~80 pixels), the output pulse is focused using phase-conjugation of the time-gated TM. Our scattering sample is too thick to exhibit memory effect~\cite{vellekoop2010scattered}, we thus cannot scan the focus by tilting the input wavefront. A similar protocol using the BBTM on a thinner SHG sample was proposed in~\cite{de_aguiar_enhanced_2016}.
After verifying that the output pulse is well focused on the CCD camera, we measure the corresponding 2PEF signal on the EMCCD camera. We use the spatially-integrated 2PEF signal for the reconstruction image at this specific focus position, to simulate an integrating detector such as a photomultiplier. The protocol is then iterated for all the $N_{\text{CCD}}$ pixels. 
Reconstruction of the 2PEF image, via this point-scanning imaging protocol, is shown in Fig.~\ref{fig:2PEF_imaging}a. We clearly identify four isolated fluorescent microbeads. Fluorescent intensities  of each microbead are different from each other, as it depends on both the fluorescent properties of each bead and the linear focus SBR which is not rigorously identical in all the positions. The measurement time of the image depends on the field of view, and the exposure time to measure the 2PEF signal at a given position. In this experiment, the lengthy measurement time ($\sim$~15h) is mostly limited by the detection process of the 2PEF signal ($T_e^{\text{EMCCD}}=$~10 s per output pixel) and not by the SLM. However it could be considerably reduced using a photomultiplier tube to collect the fluorescent signal instead of an EMCCD camera: the measurement time would then ultimately be limited by the refresh rate of the SLM, which could be made even faster using a faster SLM~\cite{blochet2017focusing}. 
The intensity profile of the lower right-hand microbead of Fig.~\ref{fig:2PEF_imaging}a, presented in Fig.~\ref{fig:2PEF_imaging}b, has a peak whose FWHM~$\sim~1~\mu$m corresponds to the microbead dimension. 
The resolution is dictated by the speckle grain size, and should be close to a widefield microscope resolution, as reported in~\cite{vellekoop2010scattered}.
In order to perform an additional check of the image, we mimic a uniform illumination by averaging the transmitted light over different illuminations on the SLM. We display 100 different random patterns on the SLM, and we measure the corresponding speckle pattern on the CCD camera. The mean linear image is shown in Figure~\ref{fig:2PEF_imaging}c. The 4 microbeads can be located thanks to diffraction effects. 

In conclusion, we have studied how the transmission matrix approach of a multiple scattering medium can be extended to ultrashort pulses of light and to multiphoton fluorescence for microbeads excitation. Comparing various approaches, we have shown that the time-gated TM enables a higher 2PEF SBR of focusing than the other techniques for a thick scattering sample, and using the MSTM to demonstrate deterministic coherent control of the two-photon process with an adjustable chirp. We then exploited the time-gated TM to perform point-scanning imaging through a thick scattering medium, where multiple scattering would naturally limit the performance of a 2PEF process. This work could have potential interest in non-linear imaging through biological tissues to enhance the SBR~\cite{ouzounov2017vivo}, notably by using the BBTM for thin multiple scattering sample~\cite{Mounaix:17,de_aguiar_enhanced_2016}, together with a fast modulation system~\cite{blochet2017focusing}.

\paragraph{Funding Information}
This work was funded by the European Research Council (COMEDIA Grant No. 278025 and SMARTIES Grant No. 724473). S.G. is a member of the Institut Universitaire de France.  

\paragraph{Acknowledgments}

The authors thank Hilton B. de Aguiar and Baptiste Blochet for fruitful discussions.

\bibliographystyle{apsrev4-1}
\bibliography{biblio_2PEF}

\end{document}